\documentclass[a4paper,11pt, svgnames]{article}
\usepackage{pos}

\usepackage{siunitx}
\usepackage{physics}
\usepackage{todonotes}

\usepackage{bbm}
\usepackage{overpic}

\usepackage{enumitem}

\usepackage{csquotes} 

\usepackage{subcaption}
\usepackage{graphicx}

\usepackage{pgf}
\usepackage{tikz, tikz-3dplot}

\usetikzlibrary{trees}
\usetikzlibrary{decorations.pathmorphing}
\usetikzlibrary{decorations.markings}
\usetikzlibrary{decorations.text}
\usetikzlibrary{calc}
\usetikzlibrary{positioning}
\usetikzlibrary{decorations.pathreplacing}

\usetikzlibrary{matrix,positioning}
\usetikzlibrary{3d}

\title{Comparison and efficiency of GPU accelerated optical light propagation in CORSIKA~8}
\author*[a]{Dominik Baack}
\author[a]{Jean-Marco Alameddine}

\affiliation[a]{Technical University Dortmund,\\
  Otto-Hahn-Straße 4a, Dortmund, Germany}

\onbehalf{for the CORSIKA8 collaboration} 

\emailAdd{dominik.baack@tu-dortmund.de}
\emailAdd{jeanmarco.alameddine@tu-dortmund.de}

\abstract{AI accelerators have proliferated in data centers in recent years and are now almost ubiquitous. In addition, their computational power and, most importantly, their energy efficiency are up to orders of magnitude higher than that of traditional computing. Over the last years, various methods and optimizations have been tested to use these hybrid systems for simulations in the context of astroparticle physics utilizing CORSIKA. The main focus of this talk is the propagation of optical, i.e. fluorescence and Cherenkov, photons through low density inhomogeneous media in the context of the next generation CORSIKA8 simulation framework. Different techniques used and approximations, e.g. the atmospheric model, tested during the development will be presented. The trade-off between performance and precision allows the experiment to achieve its physical precision limited to the real resolution of the experiment and not invest power and time in vanishing precision gains. The additional comparison of classical CPU-based simulations with the new methods validates these methods and allows evaluation against a known baseline. }

\ConferenceLogo{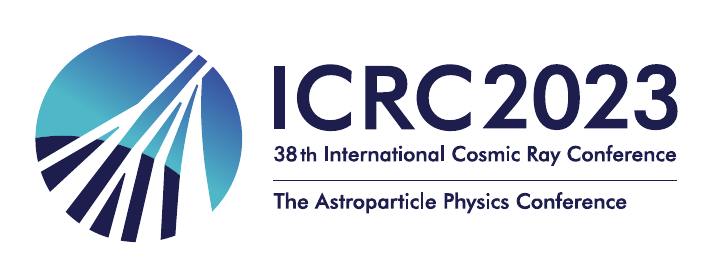}

\FullConference{%
38th International Cosmic Ray Conference (ICRC2023)\\
  26 July - 3 August, 2023\\
  Nagoya, Japan}

\begin{document}
\maketitle

\textbf{Cherenkov \& Fluorescence light simulation in EAS}
With atmospheric light propagation as one of the main contributors to runtime for IACT experiment, in the development of Corsika 8 special care was taken to accelerate this part specifically through the utilization of new or improved methods and, if available, hybrid computing.
With the extensive progress in the development of Corsika~8 Framework \cite{corsika8} over the last month, it becomes now possible to make realistic comparisons of absolute results between this framework and the present well-established Corsika 7 \cite{Heck:1998vt} simulation. 
As the overall baseline, the Corsika 7 IACT extension, developed by Conrad Bernlöhr \cite{Bernlohr_2008}, can be used. It is employed in several large-scale experiments and therefore tested against a variety of real measurements.

The redesign of Corsika was started in mid-2018, focusing on modernizing and extending the functionality. For this, rewriting in a modern programming language was necessary. Utilizing design patterns, a modular and flexible framework was created which allows the use of existing or individual modules or processes to modify the cascade development according to the user's wishes. This flexible nature allowed a direct implementation of the whole CPU-based Cherenkov queue directly into Corsika8 and is therefore a part of the core framework. 
For the utilized ICRC2023 pre-alpha Version, most parts of the code still reside in a branch, with the plan to merge it by the end of 2023. 
The GPU functionality in itself is, as of now, not part of Corsika8. The main reason behind this is the currently negligible support of parallelization in the base simulation, which is heavily required. With the GPU being several times faster than any CPU implementation, multiple parallel-running Corsika instances are required to keep up photon production to saturation. With the high costs of GPUs, this is an essential requirement. Therefore, a ZeroMQ (\O{}MQ) \cite{hintjens2011} synchronization layer, capable of using several high-speed transfer modes, was implemented and will be merged together with the Cherenkov branch. The resulting COAST-like interface can not only be useful for the here described application but, with the C++ and Python hybrid package used for the GPU-based processing as a basis, provide an extensible interface for a wide range of applications and Hardware.

\section{Light Production}
The two dominant optical light-producing processes are Cherenkov and Fluorescence emission, both differ in their underlying physical processes and in their mathematical handling significantly, which results in independent treatment and acceleration methods. As a result, only small parts of the code, such as data definitions, can be shared, which ultimately led to the two implementations being fundamentally separated from each other. Only after the light generation, the individual photon data is written to a common memory block to be processed further downstream of the processing pipeline. The described methods are optimized to run on CPU as well as on GPU-accelerated hardware with good performance.

\subsection{Fluorescence Light}
The isotropic Fluorescence emission is currently implemented in a simplified manner, which allowed for testing of the methods involved, but does not depict the physical reality with enough accuracy for experimental comparisons. As a basis, the AIRFLY parametrization \cite{AVE200741} was used:
\begin{equation}
\dv{N}{x} = Y_{\lambda} \left( P,T,h \right) \cdot \dv{E}{x}
\end{equation}
where the Yield $Y_{\lambda}$ for each specific wavelength normally depends on pressure $P$, temperature $T$ and humidity $h$. To cut down on complexity here, all dependencies were removed, and the Yield was defined as a simple constant that should match an average atmosphere, height, and emission wavelength between \SI{340}{\nano\meter} to \SI{450}{\nano\meter}.
As for ionization $\dv{E}{x}$ it was replaced by the total energy loss over the whole track, for electrons given by the electromagnetic interaction model used in CORSIKA 8 (PROPOSAL \cite{Dunsch:2018nsc}). The resulting number of photon emission positions is randomly distributed along the track. 

\subsubsection{Optimization}
Independent of how the actual number of photons is calculated, one can decrease it by reducing the normally isotropic emission to be beamed in the direction of the telescopes. For this, an opening angle and pointing depending on the telescope distance can be calculated. From the resulting solid angle, the required scaling factor is calculated and applied to the total number of photons.
Through the comparable low number of photons generated for each track, it is vital that the above method must be very fast, otherwise trivially generating all photons and applying a cut afterward will be faster. The main complexity where this method breaks down is as soon as multiple telescopes are involved. Here the process of overlapping the region of the solid emission angle is required to calculate the number of photons correctly or the telescopes needed to be handled iteratively, both methods significantly increase the number of calculations. Therefore, for multiple telescopes, an emission area is defined which includes all telescopes involved. The average gain for ground base telescopes is still better than \SI{50}{\percent} in the worst case, and therefore still acceptable compared to the additional complexity.

\subsection{Cherenkov Light}
Cherenkov photons are generated from individual particles tracks, where the number of photons is calculated via the Frank-Tamm Formula:
\begin{align}
N &=  \mu_r \pi z^2 \frac{e^2}{h \varepsilon_0 c_0} \int_{\lambda_1}^{\lambda_2} \int_{l_1}^{l_2} \frac{1}{\lambda^2}  \left( 1-\frac{1}{\beta^2 n^2(\lambda, l)}\right) \Theta \left(\beta \left(l\right) - \frac{1}{n\left(\lambda, l\right) } \right)  \dd{\lambda} \dd{l} \nonumber 
\end{align}
To handle the magnetic field and multiple scattering with enough precision, the maximum step length that the particle is allowed to travel is limited to a couple of meters. For these short travel distances, the Atmosphere is estimated as homogenous and position dependence of the refractive index $n\left(\lambda, l\right)$ is discarded.

\begin{figure}[h]
    \centering
    \begin{tabular}[t]{c|c}        
        \begin{subfigure}[c]{0.5\textwidth}
            \centering
            \begin{tikzpicture}[dot/.style={draw,fill,circle,inner sep=0pt}]
    \def\a{2cm} % large half axis
    \def\b{0.5cm} % small half axis
    \draw[line width=0.6mm, Lime] (-90:2.0cm)++(0, 3.747cm) arc (-90:-63.0:2.0cm)++(0, 3.747cm);
    \node[Lime, anchor=west, yshift=3.747cm] at  (-63:2.0cm) {$\Theta_{\text{c}}$}; 
    \fill[top color=DeepSkyBlue!90!White, bottom color=DeepSkyBlue!70!White, middle color=DeepSkyBlue!20!Black, shading=axis, opacity=0.25] (0,0) circle ({\a} and {\b});
    \fill[left color=DeepSkyBlue!60, right color=DeepSkyBlue!60, middle color=DeepSkyBlue!50, shading=axis, opacity=0.25] ({\a},0) -- (0,4) -- ({-\a},0) arc (180:360:2cm and 0.5cm);
    \draw (-2,0) arc (180:360:2cm and 0.5cm) -- (0,4) -- cycle;
    \draw[densely dashed] (-2,0) arc (180:0:2cm and 0.5cm);
    \node (X1) at (-120:{\a} and {\b}) {}; 
    \node (X2) at (-135:{\a} and {\b}) {};
    \node (X3) at (-13:{\a} and {\b}) {};
    \node (X4) at (-123:{\a} and {\b}) {};
    \node (X5) at (-76:{\a} and {\b}) {};
    \node (X6) at (-145:{\a} and {\b}) {};
    \node (X7) at (-106:{\a} and {\b}) {};
    \node (X8) at (-14:{\a} and {\b}) {};
    \node (X9) at (-69:{\a} and {\b}) {};
    \node (X10) at (-150:{\a} and {\b}) {};
    \node (X11) at (-14:{\a} and {\b}) {};

    \node (X21) at (68:{\a} and {\b}) {};
    \node (X22) at (23:{\a} and {\b}) {};
    \node (X23) at (114:{\a} and {\b}) {};
    \node (X24) at (169:{\a} and {\b}) {};
    \node (X25) at (1:{\a} and {\b}) {};
    \node (X26) at (51:{\a} and {\b}) {};
    \node (X27) at (33:{\a} and {\b}) {};
    \node (X28) at (82:{\a} and {\b}) {};
    \node (X29) at (127:{\a} and {\b}) {};

    \draw[color=Blue, opacity=0.75, -stealth, outer sep=0] (0,4cm) -- (X1);
    \draw[color=Blue, opacity=0.75, -stealth, outer sep=0] (0,4cm) -- (X2);
    \draw[color=Blue, opacity=0.75, -stealth, outer sep=0] (0,4cm) -- (X3);
    \draw[color=Blue, opacity=0.75, -stealth, outer sep=0] (0,4cm) -- (X4);
    \draw[color=Blue, opacity=0.75, -stealth, outer sep=0] (0,4cm) -- (X5);
    \draw[color=Blue, opacity=0.75, -stealth, outer sep=0] (0,4cm) -- (X6);
    \draw[color=Blue, opacity=0.75, -stealth, outer sep=0] (0,4cm) -- (X7);
    \draw[color=Blue, opacity=0.75, -stealth, outer sep=0] (0,4cm) -- (X8);
    \draw[color=Blue, opacity=0.75, -stealth, outer sep=0] (0,4cm) -- (X9);
    \draw[color=Blue, opacity=0.75, -stealth, outer sep=0] (0,4cm) -- (X10);
    \draw[color=Blue, opacity=0.75, -stealth, outer sep=0] (0,4cm) -- (X11);
    \draw[color=Blue!70, opacity=0.75, -stealth, outer sep=0] (0,4cm) -- (X21);
    \draw[color=Blue!70, opacity=0.75, -stealth, outer sep=0] (0,4cm) -- (X22);
    \draw[color=Blue!70, opacity=0.75, -stealth, outer sep=0] (0,4cm) -- (X23);
    \draw[color=Blue!70, opacity=0.75, -stealth, outer sep=0] (0,4cm) -- (X24);
    \draw[color=Blue!70, opacity=0.75, -stealth, outer sep=0] (0,4cm) -- (X25);
    \draw[color=Blue!70, opacity=0.75, -stealth, outer sep=0] (0,4cm) -- (X26);
    \draw[color=Blue!70, opacity=0.75, -stealth, outer sep=0] (0,4cm) -- (X27);
    \draw[color=Blue!70, opacity=0.75, -stealth, outer sep=0] (0,4cm) -- (X28);
    \draw[color=Blue!70, opacity=0.75, -stealth, outer sep=0] (0,4cm) -- (X29);
   
    %
    % Z Axis   
    \draw[line width=0.2mm, black] (0, -1) -- (0, 5.0);
    \draw[line width=0.2mm, black] (-0.1, 4.8) -- (0.1, 5.2);
    \draw[line width=0.2mm, black] (-0.1, 4.9) -- (0.1, 5.3);
    \draw[line width=0.2mm, black, -latex] (0, 5.1) -- (0, 6) node[anchor=south]{Z};
    %
    % X Axis
    \draw[line width=0.2mm, black] (0, 4) -- (0.8, 4);
    \draw[line width=0.2mm, black] (0.6, 4.1) -- (1.0, 3.9);
    \draw[line width=0.2mm, black] (0.7, 4.1) -- (1.1, 3.9);
    \draw[line width=0.2mm, black, -latex] (0.9, 4) -- (2, 4) node[anchor=west]{X};
    %
    % Y Axis
    \draw[line width=0.2mm, black,] (0, 4) -- (0.4, 4.4);
    \draw[line width=0.2mm, black,] ({0.4 - 3*0.07071}, {4.4 - 1*0.07071}) -- ({0.4 + 3*0.07071}, {4.4 + 1*0.07071});
    \draw[line width=0.2mm, black,] ({0.4 - 3*0.07071 + 0.07071}, {4.4 - 1*0.07071 + 0.07071}) -- ({0.4 + 3*0.07071 + 0.07071}, {4.4 + 1*0.07071 + 0.07071});
    \draw[line width=0.2mm, black,] (0, 4) -- (0.4, 4.4);
    \draw[line width=0.2mm, black, -latex] (0.47071, 4.47071) -- (1, 5) node[anchor=south west]{Y};  
    \draw[line width=1.0mm, red, -latex] (0:{\a} and {\b}) arc (0:-120:{\a} and {\b});
    \node[red, anchor=north] at  (-55:{\a} and {\b}) {$c_{\text{rng}}$};   
    \draw[black, fill=black] (-1, 7) circle (0.05cm) node [anchor=west]{Particle};
    \draw[line width=0.2mm, black, -latex] (-1, 7) -- (-2, 6);  
\end{tikzpicture}
            \caption{Photon generation on a cone}    
            \label{fig:cherenkov_generation_a}    
        \end{subfigure}%
    &
    \begin{tabular}{c}% if you add [t], than sub images are pushed down
        \smallskip
        \begin{subfigure}[t]{0.4\textwidth}
            \centering
            \begin{tikzpicture}[dot/.style={draw,fill,circle,inner sep=0pt}]   
    % Z Axis   
    \draw[line width=0.2mm, black] (0, 0) -- (0, 1.0);
    \draw[line width=0.2mm, black] (-0.1, 0.8) -- (0.1, 1.2);
    \draw[line width=0.2mm, black] (-0.1, 0.9) -- (0.1, 1.3);
    \draw[line width=0.2mm, black, -latex] (0, 0.1) -- (0, 2) node[anchor=south]{Z};
    %
    % X Axis
    \draw[line width=0.2mm, black] (0, 0) -- (1.0, 0);
    \draw[line width=0.2mm, black] (0.8, 0.1) -- (1.2, -0.1);
    \draw[line width=0.2mm, black] (0.9, 0.1) -- (1.3, -0.1);
    \draw[line width=0.2mm, black, -latex] (1.1, 0) -- (2, 0) node[anchor=west]{X};
    %
    % Y Axis
    \draw[line width=0.2mm, black,] (0, 0) -- (0.7071, 0.7071);
    \draw[line width=0.2mm, black,] ({0.7071 - 3*0.07071}, {0.7071 - 1*0.07071}) -- ({0.7071 + 3*0.07071}, {0.7071 + 1*0.07071});
    \draw[line width=0.2mm, black,] ({0.7071 - 3*0.07071 + 0.07071}, {0.7071 - 1*0.07071 + 0.07071}) -- ({0.7071 + 3*0.07071 + 0.07071}, {0.7071 + 1*0.07071 + 0.07071});
    \draw[line width=0.2mm, black, -latex] ({0.7071+ 0.07071}, {0.7071+ 0.07071}) -- (1.4142, 1.4142) node[anchor=south west]{Y};  
    \draw[black, fill=black] (-1, 3) circle (0.05cm) node [anchor=west]{Particle};
    \draw[line width=0.2mm, black, -latex] (-1, 3) -- (-2, 2);

    \begin{scope}[yshift = -0.84cm, xshift=-0.84cm, scale=0.2, rotate around={-45:(0,0)}]
        \def\a{2cm} % large half axis
        \def\b{0.5cm} % small half axis
        \fill[top color=DeepSkyBlue!90!White, bottom color=DeepSkyBlue!70!White, middle color=DeepSkyBlue!20!Black, shading=axis, opacity=0.25] (0,0) circle ({\a} and {\b});
        \fill[left color=DeepSkyBlue!60, right color=DeepSkyBlue!60, middle color=DeepSkyBlue!50, shading=axis, opacity=0.25] ({\a},0) -- (0,6) -- ({-\a},0) arc (180:360:2cm and 0.5cm);
        \draw (-2,0) arc (180:360:2cm and 0.5cm) -- (0,6) -- cycle;
        \draw[densely dashed] (-2,0) arc (180:0:2cm and 0.5cm);
        \node (X1) at (-120:{\a} and {\b}) {}; 
        \node (X2) at (-135:{\a} and {\b}) {};
        \node (X3) at (-13:{\a} and {\b}) {};
        \node (X4) at (-123:{\a} and {\b}) {};
        \node (X5) at (-76:{\a} and {\b}) {};
        \node (X6) at (-145:{\a} and {\b}) {};
        \node (X7) at (-106:{\a} and {\b}) {};
        \node (X8) at (-14:{\a} and {\b}) {};
        \node (X9) at (-69:{\a} and {\b}) {};
        \node (X10) at (-150:{\a} and {\b}) {};
        \node (X11) at (-14:{\a} and {\b}) {};
        
        \node (X21) at (68:{\a} and {\b}) {};
        \node (X22) at (23:{\a} and {\b}) {};
        \node (X23) at (114:{\a} and {\b}) {};
        \node (X24) at (169:{\a} and {\b}) {};
        \node (X25) at (1:{\a} and {\b}) {};
        \node (X26) at (51:{\a} and {\b}) {};
        \node (X27) at (33:{\a} and {\b}) {};
        \node (X28) at (82:{\a} and {\b}) {};
        \node (X29) at (127:{\a} and {\b}) {};

        \draw[color=Blue, opacity=0.75, -stealth, outer sep=0] (0,6cm) -- (X1);
        \draw[color=Blue, opacity=0.75, -stealth, outer sep=0] (0,6cm) -- (X2);
        \draw[color=Blue, opacity=0.75, -stealth, outer sep=0] (0,6cm) -- (X3);
        \draw[color=Blue, opacity=0.75, -stealth, outer sep=0] (0,6cm) -- (X4);
        \draw[color=Blue, opacity=0.75, -stealth, outer sep=0] (0,6cm) -- (X5);
        \draw[color=Blue, opacity=0.75, -stealth, outer sep=0] (0,6cm) -- (X6);
        \draw[color=Blue, opacity=0.75, -stealth, outer sep=0] (0,6cm) -- (X7);
        \draw[color=Blue, opacity=0.75, -stealth, outer sep=0] (0,6cm) -- (X8);
        \draw[color=Blue, opacity=0.75, -stealth, outer sep=0] (0,6cm) -- (X9);
        \draw[color=Blue, opacity=0.75, -stealth, outer sep=0] (0,6cm) -- (X10);
        \draw[color=Blue, opacity=0.75, -stealth, outer sep=0] (0,6cm) -- (X11);
        \draw[color=Blue!70, opacity=0.75, -stealth, outer sep=0] (0,6cm) -- (X21);
        \draw[color=Blue!70, opacity=0.75, -stealth, outer sep=0] (0,6cm) -- (X22);
        \draw[color=Blue!70, opacity=0.75, -stealth, outer sep=0] (0,6cm) -- (X23);
        \draw[color=Blue!70, opacity=0.75, -stealth, outer sep=0] (0,6cm) -- (X24);
        \draw[color=Blue!70, opacity=0.75, -stealth, outer sep=0] (0,6cm) -- (X25);
        \draw[color=Blue!70, opacity=0.75, -stealth, outer sep=0] (0,6cm) -- (X26);
        \draw[color=Blue!70, opacity=0.75, -stealth, outer sep=0] (0,6cm) -- (X27);
        \draw[color=Blue!70, opacity=0.75, -stealth, outer sep=0] (0,6cm) -- (X28);
        \draw[color=Blue!70, opacity=0.75, -stealth, outer sep=0] (0,6cm) -- (X29);   
    \end{scope}
\end{tikzpicture}
            \caption{Rotate into particle frame.}        
        \end{subfigure}\\
        \begin{subfigure}[t]{0.4\textwidth}
            \centering
            \begin{tikzpicture}[dot/.style={draw,fill,circle,inner sep=0pt}]   
    % Z Axis   
    \draw[line width=0.2mm, black] (0, 0) -- (0, 1.0);
    \draw[line width=0.2mm, black] (-0.1, 0.8) -- (0.1, 1.2);
    \draw[line width=0.2mm, black] (-0.1, 0.9) -- (0.1, 1.3);
    \draw[line width=0.2mm, black, -latex] (0, 0.1) -- (0, 2) node[anchor=south]{Z};
    %
    % X Axis
    \draw[line width=0.2mm, black] (0, 0) -- (1.0, 0);
    \draw[line width=0.2mm, black] (0.8, 0.1) -- (1.2, -0.1);
    \draw[line width=0.2mm, black] (0.9, 0.1) -- (1.3, -0.1);
    \draw[line width=0.2mm, black, -latex] (1.1, 0) -- (2, 0) node[anchor=west]{X};
    %
    % Y Axis
    \draw[line width=0.2mm, black,] (0, 0) -- (0.7071, 0.7071);
    \draw[line width=0.2mm, black,] ({0.7071 - 3*0.07071}, {0.7071 - 1*0.07071}) -- ({0.7071 + 3*0.07071}, {0.7071 + 1*0.07071});
    \draw[line width=0.2mm, black,] ({0.7071 - 3*0.07071 + 0.07071}, {0.7071 - 1*0.07071 + 0.07071}) -- ({0.7071 + 3*0.07071 + 0.07071}, {0.7071 + 1*0.07071 + 0.07071});
    \draw[line width=0.2mm, black, -latex] ({0.7071+ 0.07071}, {0.7071+ 0.07071}) -- (1.4142, 1.4142) node[anchor=south west]{Y};  
    \draw[black, fill=black] (-1, 3) circle (0.05cm) node [anchor=west]{Particle};
    \draw[line width=0.2mm, black, -latex] (-1, 3) -- (-2, 2);

    \begin{scope}[yshift = 1.75cm, xshift=-2.25cm, rotate around={-45:(0,0)}, scale=0.2]
        \def\a{2cm} % large half axis
        \def\b{0.5cm} % small half axis
        \fill[top color=DeepSkyBlue!90!White, bottom color=DeepSkyBlue!70!White, middle color=DeepSkyBlue!20!Black, shading=axis, opacity=0.25] (0,0) circle ({\a} and {\b});
        \fill[left color=DeepSkyBlue!60, right color=DeepSkyBlue!60, middle color=DeepSkyBlue!50, shading=axis, opacity=0.25] ({\a},0) -- (0,6) -- ({-\a},0) arc (180:360:2cm and 0.5cm);
        \draw (-2,0) arc (180:360:2cm and 0.5cm) -- (0,6) -- cycle;
        \draw[densely dashed] (-2,0) arc (180:0:2cm and 0.5cm);
        \node (X1) at (-120:{\a} and {\b}) {}; 
        \node (X2) at (-135:{\a} and {\b}) {};
        \node (X3) at (-13:{\a} and {\b}) {};
        \node (X4) at (-123:{\a} and {\b}) {};
        \node (X5) at (-76:{\a} and {\b}) {};
        \node (X6) at (-145:{\a} and {\b}) {};
        \node (X7) at (-106:{\a} and {\b}) {};
        \node (X8) at (-14:{\a} and {\b}) {};
        \node (X9) at (-69:{\a} and {\b}) {};
        \node (X10) at (-150:{\a} and {\b}) {};
        \node (X11) at (-14:{\a} and {\b}) {};
        
        \node (X21) at (68:{\a} and {\b}) {};
        \node (X22) at (23:{\a} and {\b}) {};
        \node (X23) at (114:{\a} and {\b}) {};
        \node (X24) at (169:{\a} and {\b}) {};
        \node (X25) at (1:{\a} and {\b}) {};
        \node (X26) at (51:{\a} and {\b}) {};
        \node (X27) at (33:{\a} and {\b}) {};
        \node (X28) at (82:{\a} and {\b}) {};
        \node (X29) at (127:{\a} and {\b}) {};

        \draw[color=Blue, opacity=0.75, -stealth, outer sep=0] (0,6cm) -- (X1);
        \draw[color=Blue, opacity=0.75, -stealth, outer sep=0] (0,6cm) -- (X2);
        \draw[color=Blue, opacity=0.75, -stealth, outer sep=0] (0,6cm) -- (X3);
        \draw[color=Blue, opacity=0.75, -stealth, outer sep=0] (0,6cm) -- (X4);
        \draw[color=Blue, opacity=0.75, -stealth, outer sep=0] (0,6cm) -- (X5);
        \draw[color=Blue, opacity=0.75, -stealth, outer sep=0] (0,6cm) -- (X6);
        \draw[color=Blue, opacity=0.75, -stealth, outer sep=0] (0,6cm) -- (X7);
        \draw[color=Blue, opacity=0.75, -stealth, outer sep=0] (0,6cm) -- (X8);
        \draw[color=Blue, opacity=0.75, -stealth, outer sep=0] (0,6cm) -- (X9);
        \draw[color=Blue, opacity=0.75, -stealth, outer sep=0] (0,6cm) -- (X10);
        \draw[color=Blue, opacity=0.75, -stealth, outer sep=0] (0,6cm) -- (X11);
        \draw[color=Blue!70, opacity=0.75, -stealth, outer sep=0] (0,6cm) -- (X21);
        \draw[color=Blue!70, opacity=0.75, -stealth, outer sep=0] (0,6cm) -- (X22);
        \draw[color=Blue!70, opacity=0.75, -stealth, outer sep=0] (0,6cm) -- (X23);
        \draw[color=Blue!70, opacity=0.75, -stealth, outer sep=0] (0,6cm) -- (X24);
        \draw[color=Blue!70, opacity=0.75, -stealth, outer sep=0] (0,6cm) -- (X25);
        \draw[color=Blue!70, opacity=0.75, -stealth, outer sep=0] (0,6cm) -- (X26);
        \draw[color=Blue!70, opacity=0.75, -stealth, outer sep=0] (0,6cm) -- (X27);
        \draw[color=Blue!70, opacity=0.75, -stealth, outer sep=0] (0,6cm) -- (X28);
        \draw[color=Blue!70, opacity=0.75, -stealth, outer sep=0] (0,6cm) -- (X29);   
    \end{scope}
\end{tikzpicture}
            \caption{Translation to particle track.}        
        \end{subfigure}
        \end{tabular}\\    
    \end{tabular}
    \caption{Schematic representation of the individual steps for the Cherenkov photon generation.}
    \label{fig:cherenkov_generation_schema}
\end{figure}
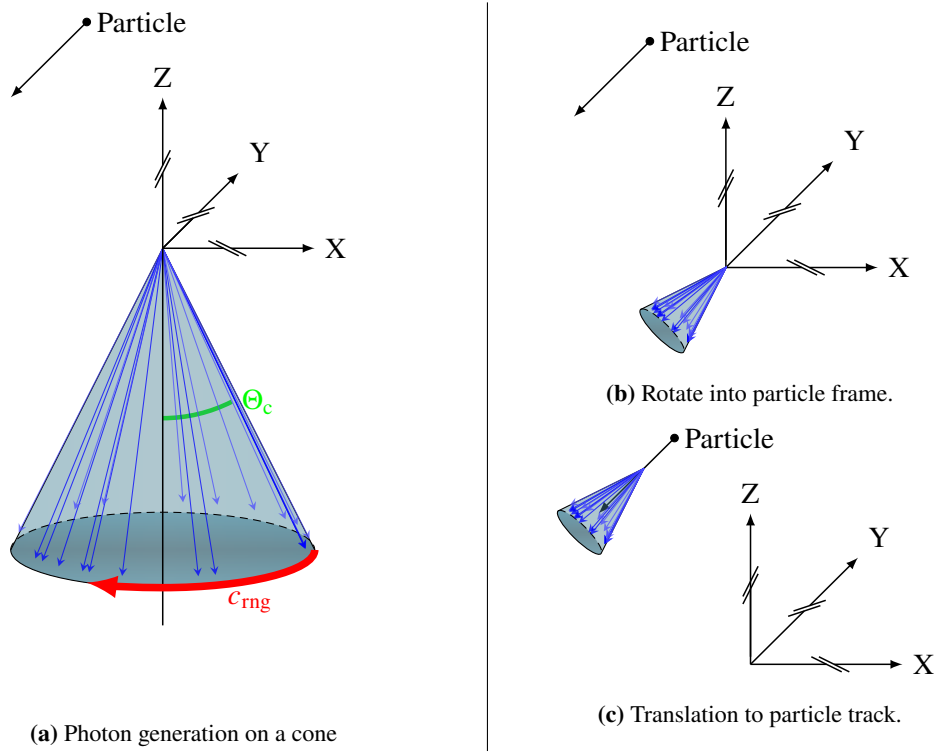

The photons themselves are generated in the downward-facing direction in the coordinate system's origin, with the correct Cherenkov angle. The photon cone is then rotated into the pointing direction of the particle track before they get moved to its corresponding emission point. The individual steps are schematically displayed in figure \ref{fig:cherenkov_generation_schema}.

\subsubsection{Optimization}
For the highly beamed Cherenkov light emission, the reduction in photon calculation is much simpler to apply compared to fluorescence. The easiest method is to calculate the angle difference between the currently handled particles' flight direction and the vector to the experiment center. For higher precision, the minimum angle to each individual telescope can be used. As long as the individual telescopes do not exceed the meter range, this method can reduce the number of handled particles by up to \SI{90}{\percent}, which results in a similar impressive speedup. For larger-sized telescopes, the method requires a distance surpassing \SI{\approx 15}{} times the radius to work with the recommended cut of \SI{6}{\degree} without visible photon loss. For telescopes exceeding \SI{50}{\meter} in diameter, the instrumented area should be split into several smaller-sized domains or an algorithm used which does not approximate the telescope as a singular point.

\begin{figure}
     \begin{subfigure}[c]{0.5\textwidth}
        \centering
        \begin{overpic}[width=\textwidth]{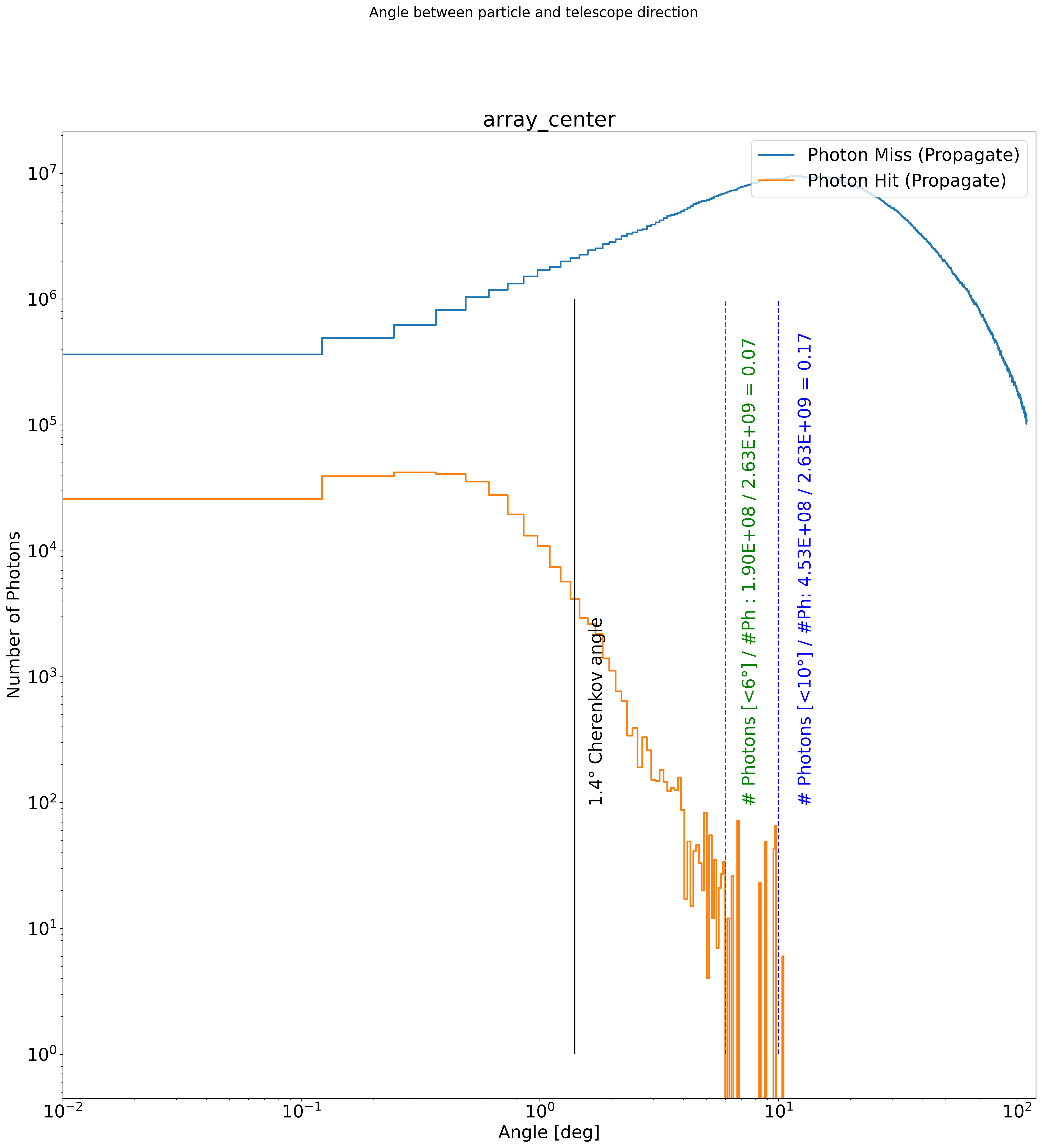}          
            \put(-2,-1){\includegraphics[width=2.2cm]{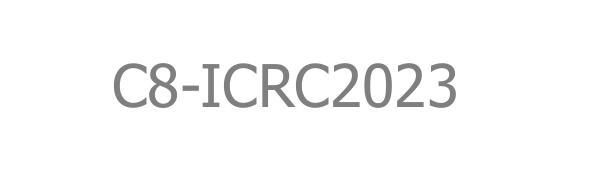}}
        \end{overpic}
        \caption{Angle to center of whole telescope array}    
        \label{fig:angle_dist_array}    
    \end{subfigure}%
    \begin{subfigure}[c]{0.5\textwidth}
        \centering
        \begin{overpic}[width=\textwidth]{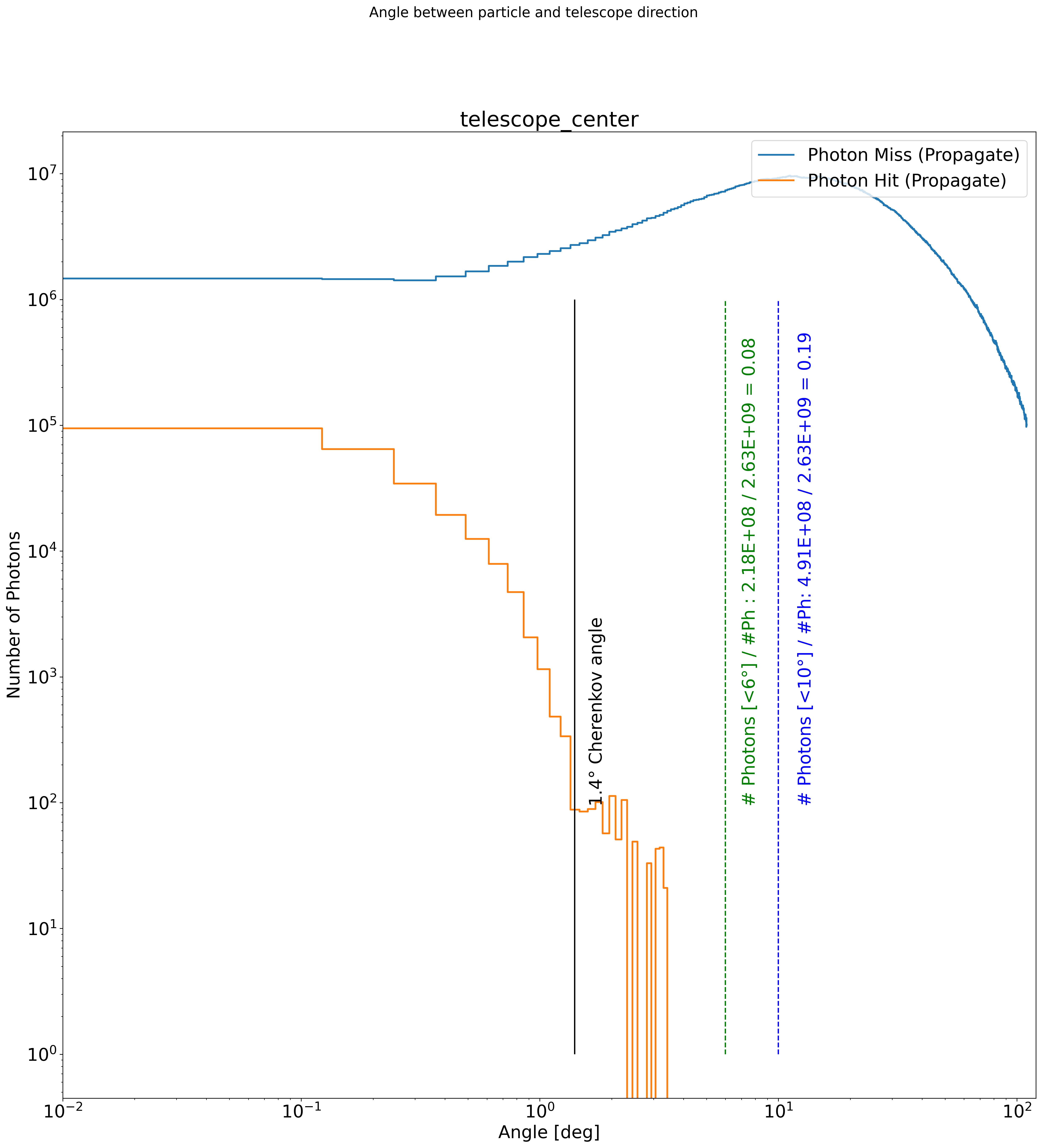}            
            \put(-2,-1){\includegraphics[width=2.2cm]{img/watermark.pdf}}
        \end{overpic}         
        \caption{Angle to center of individual telescopes}    
        \label{fig:angle_dist_telescope}    
    \end{subfigure}%   
    \caption{Efficiency of an angular particle cut for the reduction of atmospheric photons.}
    \label{fig:particle_cut_cherenkov}
\end{figure}

The second small-scale optimization is the exact method of how the photon generation is implemented, specifically the rotation matrix. Here, the photon vector
\begin{align*}
    \vec{p} &= \left( \begin{pmatrix} %
        &\cos(c_{\text{rng}}) \cdot \sin(\Theta_{\text{c}}) \\
        &\sin(c_{\text{rng}}) \cdot \sin(\Theta_{\text{c}}) \\
        &-\cos(\Theta_{\text{c}})\end{pmatrix} \right)
\end{align*}%
is multiplied by the following matrix:
\begin{align*}    
    M_{\text{Rod}} &= \mathbbm{1} + 2 \cdot R_{\text{Rod}}^2 \\
    &=\begin{pmatrix}
        - 2 A_y^{2} - 2 A_z^{2} + 1 & 2 A_x A_y & 2 A_x A_z\\
        2 A_x A_y & - 2 A_x^{2} - 2 A_z^{2} + 1 & 2 A_y A_z\\
        2 A_x A_z & 2 A_y A_z & - 2 A_x^{2} - 2 A_y^{2} + 1
    \end{pmatrix}
\end{align*}
where the vector $\vec{A}$ is defined as the bisecting vector between the particle flight direction and the downward-facing normal vector.

\section{Photon Propagation}
After generating Photons at their point of origin they must be propagated to ground level which is done by the fastest method currently implemented consisting of a straight line propagation followed by the application of correction factors which changes the arrival angle, position, and timing to comply with atmospheric refraction. Details can be found in an earlier release about the Cherenkov light propagation \cite{alameddine2021corsika}.

\section{Absolute Comparison Corsika7 \& Corsika8}
To compare the new Corsika8 implementation and methods against the current Corsika7 Version, two different methods were utilized. The first is the direct comparison with absolute values, which give conclusions to the development of the whole cascade. The second method, which isolates the influence from all other modules, is replaying the Corsika7-generated cascade in Corsika8 and applying only those processes that are required for the comparison. This allows for a direct comparison, eliminating all other sources of interference.
It is to mention that even if Corsika7 is used as a reference it is still a simulation which implies the necessity of simplification and resulting discrepancy compared to real-world measurements. So slight deviations between the simulations are to be expected and do not inevitably imply an error in either simulation.
 
\subsection{Intensity \& Timing Profile}
\begin{figure}
\centering
\vspace{-2em}
    \begin{subfigure}[b]{0.7\textwidth}
        \centering
        \includegraphics[width=\textwidth]{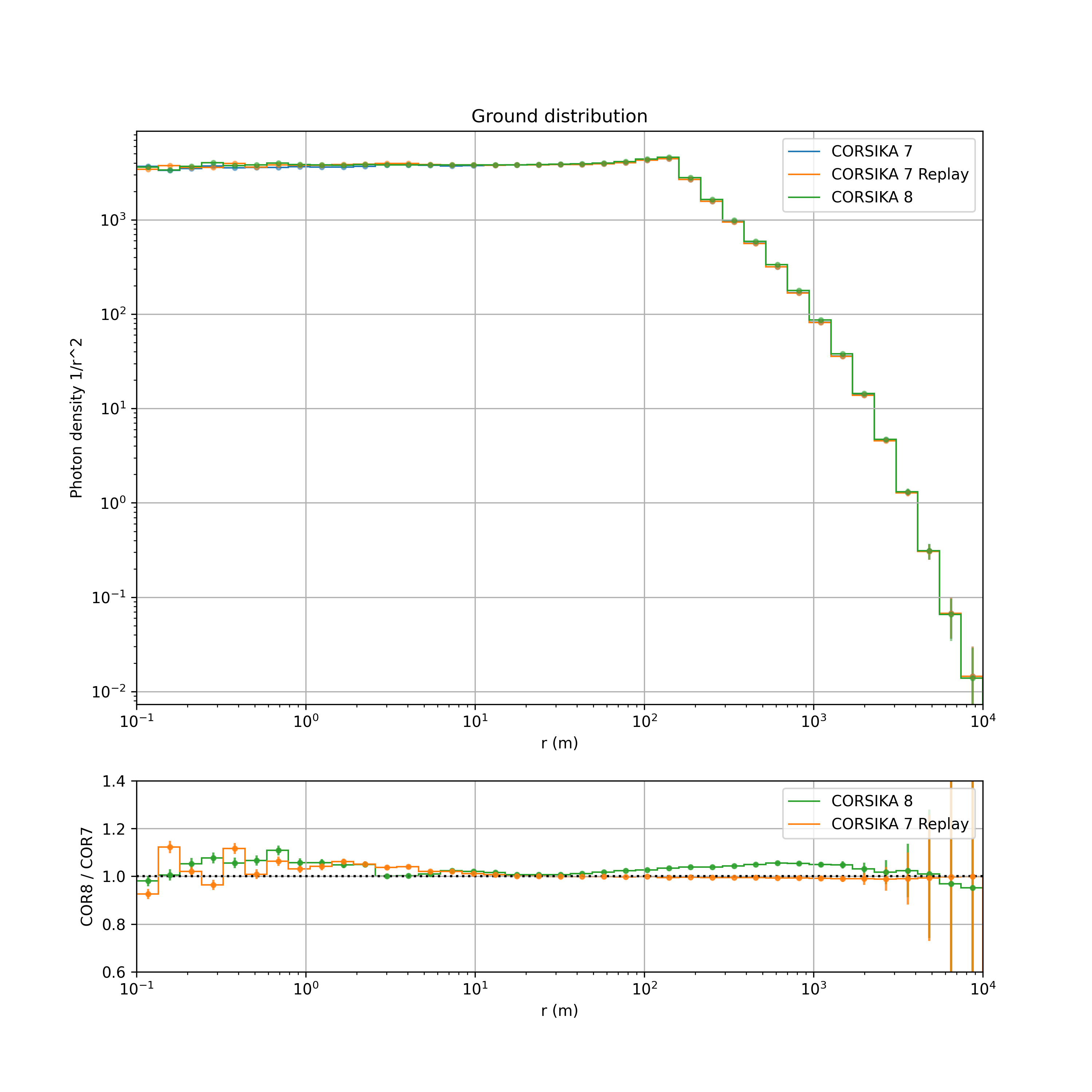}
        \caption{Displayed is the radial outgoing photon density.}
        \label{fig:radial_density}
    \end{subfigure}
    \centering
    \begin{subfigure}[b]{0.7\textwidth}
        \centering
        %trim={<left> <lower> <right> <upper>}
        \includegraphics[width=\textwidth]{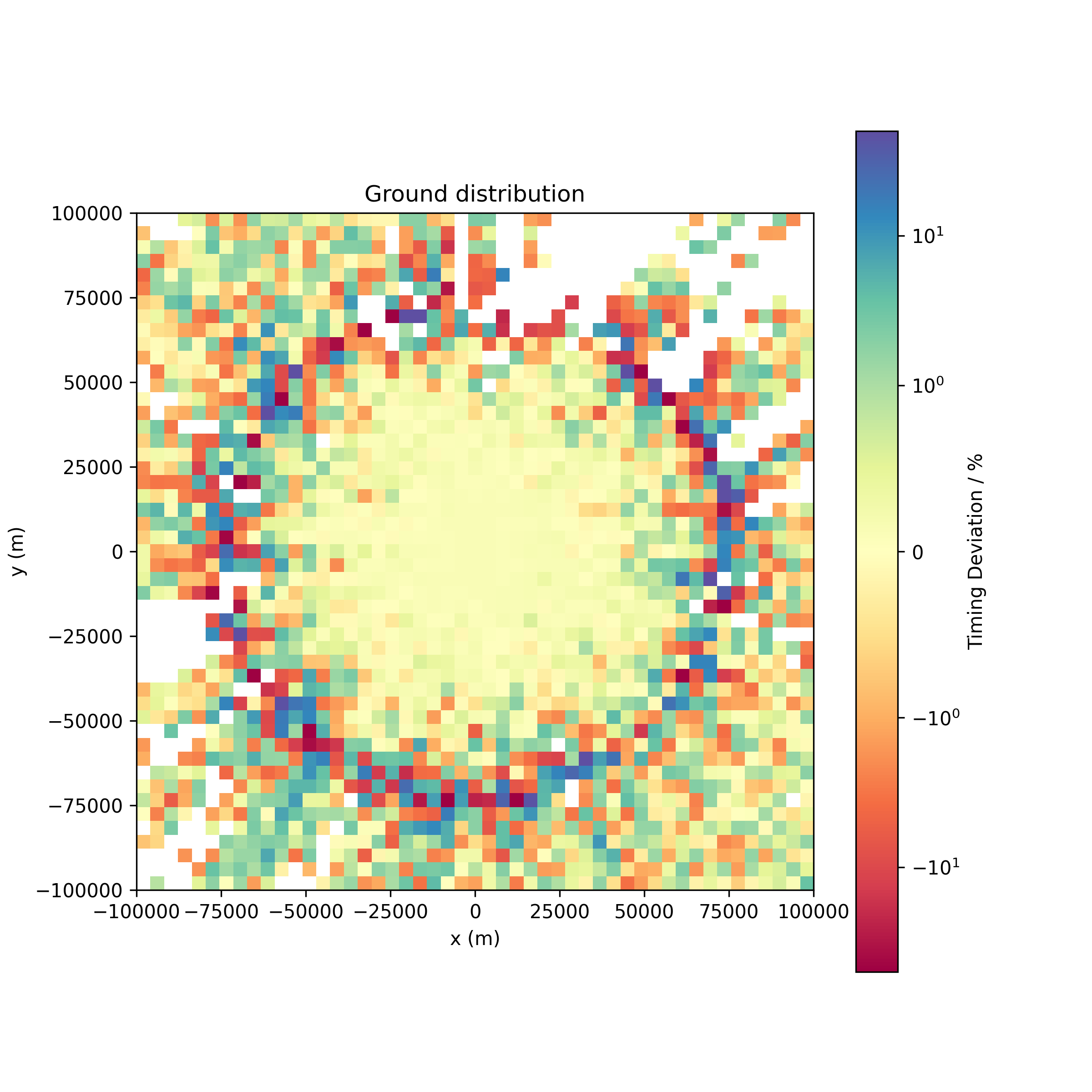}
        \caption{Displayed is the relative difference in arrival time for Corsika7 and Corsika8 normalized to the mean arrival time.}
        \label{fig:radial_timing}
    \end{subfigure}
    \caption{Distributions from Corsika7 IACT, Corsika8 and in the upper picture the newly implemented \enquote{replay} mode.}    
\end{figure}

Radial profiles from the calculated shower center are only applicable to nearly rotational symmetric ground distribution, therefore only to those simulated without inclination angle. But the resulting distribution allows good examinations between the different versions of Corsika. One such profile is displayed in figure \ref{fig:radial_density}. It shows the resulting light distribution of 100 added \SI{500}{\giga\electronvolt} Gamma-induced showers for Corsika7, Corsika8, and Corsika7 particle tracks with the Cherenkov module of Corsika8 (called replay). The result shows an exact match, therefore the photon generation and propagation produce the exact same density distribution. Because the implementation differs from the Corsika7 one, a binary match will not be possible.
The deviation between Corsika8 and Corsika7 is below \SI{10}{\percent} and explainable through the changes in interaction models.

The comparison of the mean arrival times, as displayed in figure \ref{fig:radial_timing}, exhibits similar good results. The central region, up to \SI{2000}{\meter} from the core, where enough statistics is present, has nearly \SI{0}{\percent} deviation between Corsika7 and Corsika8. The outer region is dominated by random processes of individual particles, where no decision can be made even with the performed cut of, at least, 50 Photons per bin.

\section{Conclusions and Outlook}
With a first tested implementation of optical light generation and propagation on CPU basis inside Corsika8 and a second GPU implementation running in a separate program synchronized via shared memory, the first steps for the use of Corsika8 in the IACT and Fluorescence community were taken. With further development of Corsika8 and already existing innovations like cross-media shower, the implemented interface enables new opportunities and methods not possible in Corsika7.

\let\oldbibliography\thebibliography
\renewcommand{\thebibliography}[1]{%
  \oldbibliography{#1}%
  \setlength{\itemsep}{1pt}%
}

{\footnotesize
\bibliographystyle{JHEP}
\bibliography{refs}

\providecommand{\href}[2]{#2}\begingroup\raggedright\begin{thebibliography}{1}

\bibitem{corsika8}
{Alves, Antonio Augusto}, {Reininghaus, Maximilian}, {Schmidt, Andr\'e}, {Prechelt, Remy}, {Ulrich, Ralf} and {for the CORSIKA 8 collaboration}, \emph{Corsika 8 - a novel high-performance computing tool for particle cascade monte carlo simulations}, \href{https://doi.org/10.1051/epjconf/202125103038}{\emph{EPJ Web Conf.} {\bfseries 251} (2021) 03038}.

\bibitem{Heck:1998vt}
D.~Heck, J.~Knapp, J.N.~Capdevielle, G.~Schatz and T.~Thouw, \emph{Corsika: A monte carlo code to simulate extensive air showers}, {\emph{Forschungszentrum Karlsruhe Report} (1998) }.

\bibitem{Bernlohr_2008}
K.~Bernlöhr, \emph{Simulation of imaging atmospheric cherenkov telescopes with {CORSIKA} and sim{\_}telarray}, \href{https://doi.org/10.1016/j.astropartphys.2008.07.009}{\emph{Astroparticle Physics} {\bfseries 30} (2008) 149}.

\bibitem{hintjens2011}
P.~Hintjens, \emph{0MQ - The Guide}, 2011.

\bibitem{AVE200741}
M.~Ave, M.~Bohacova, B.~Buonomo, N.~Busca, L.~Cazon, S.~Chemerisov et~al., \emph{Measurement of the pressure dependence of air fluorescence emission induced by electrons}, \href{https://doi.org/https://doi.org/10.1016/j.astropartphys.2007.04.006}{\emph{Astroparticle Physics} {\bfseries 28} (2007) 41}.

\bibitem{Dunsch:2018nsc}
M.~Dunsch, J.~Soedingrekso, A.~Sandrock, M.~Meier, T.~Menne and W.~Rhode, \emph{{Recent Improvements for the Lepton Propagator PROPOSAL}}, \href{https://doi.org/10.1016/j.cpc.2019.03.021}{\emph{Comput. Phys. Commun.} {\bfseries 242} (2019) 132} [\href{https://arxiv.org/abs/1809.07740}{{\ttfamily 1809.07740}}].

\bibitem{alameddine2021corsika}
J.-M.~Alameddine, J.~Albrecht, J.~Alvarez-Muniz, A.A.A.J.~au2, L.~Arrabito, D.~Baack et~al., \emph{Corsika 8 -- contributions to the 37th international cosmic ray conference in berlin germany (icrc 2021)},  2021.

\end{thebibliography}\endgroup
}

%% Full authors list (ONLY FOR COLLABORATIONS)
\clearpage
\section*{The CORSIKA 8 Collaboration}
\small

\begin{sloppypar}\noindent
% created on 2023-07-10
J.M.~Alameddine$^{1}$,
J.~Albrecht$^{1}$,
J.~Alvarez-Mu\~niz$^{2}$,
J.~Ammerman-Yebra$^{2}$,
L.~Arrabito$^{3}$,
J.~Augscheller$^{4}$,
A.A.~Alves Jr.$^{4}$,
D.~Baack$^{1}$,
K.~Bernl\"ohr$^{5}$,
M.~Bleicher$^{6}$,
A.~Coleman$^{7}$,
H.~Dembinski$^{1}$,
D.~Els\"asser$^{1}$,
R.~Engel$^{4}$,
A.~Ferrari$^{4}$,
C.~Gaudu$^{8}$,
C.~Glaser$^{7}$,
D.~Heck$^{4}$,
F.~Hu$^{9}$,
T.~Huege$^{4,10}$,
K.H.~Kampert$^{8}$,
N.~Karastathis$^{4}$,
U.A.~Latif$^{11}$,
H.~Mei$^{12}$,
L.~Nellen$^{13}$,
T.~Pierog$^{4}$,
R.~Prechelt$^{14}$,
M.~Reininghaus$^{15}$,
W.~Rhode$^{1}$,
F.~Riehn$^{16,2}$,
M.~Sackel$^{1}$,
P.~Sala$^{17}$,
P.~Sampathkumar$^{4}$,
A.~Sandrock$^{8}$,
J.~Soedingrekso$^{1}$,
R.~Ulrich$^{4}$,
D.~Xu$^{12}$,
E.~Zas$^{2}$

\end{sloppypar}

\begin{center}
\rule{0.1\columnwidth}{0.5pt}
\raisebox{-0.4ex}{\scriptsize$\bullet$}
\rule{0.1\columnwidth}{0.5pt}
\end{center}

\vspace{-1ex}
\footnotesize
% created on 2023-07-10
% needs \usepackage{enumitem}
\begin{description}[labelsep=0.2em,align=right,labelwidth=0.7em,labelindent=0em,leftmargin=2em,noitemsep]
\item[$^{1}$] Technische Universit\"at Dortmund (TU), Department of Physics, Dortmund, Germany
\item[$^{2}$] Universidade de Santiago de Compostela, Instituto Galego de F\'\i{}sica de Altas Enerx\'\i{}as (IGFAE), Santiago de Compostela, Spain
\item[$^{3}$] Laboratoire Univers et Particules de Montpellier, Universit\'e de Montpellier, Montpellier, France
\item[$^{4}$] Karlsruhe Institute of Technology (KIT), Institute for Astroparticle Physics (IAP), Karlsruhe, Germany
\item[$^{5}$] Max Planck Institute for Nuclear Physics (MPIK), Heidelberg, Germany
\item[$^{6}$] Goethe-Universit\"at Frankfurt am Main, Institut f\"ur Theoretische Physik, Frankfurt am Main, Germany
\item[$^{7}$] Uppsala University, Department of Physics and Astronomy, Uppsala, Sweden
\item[$^{8}$] Bergische Universit\"at Wuppertal, Department of Physics, Wuppertal, Germany
\item[$^{9}$] Peking University (PKU), School of Physics, Beijing, China
\item[$^{10}$] Vrije Universiteit Brussel, Astrophysical Institute, Brussels, Belgium
\item[$^{11}$] Vrije Universiteit Brussel, Dienst ELEM, Inter-University Institute for High Energies (IIHE), Brussels, Belgium
\item[$^{12}$] Tsung-Dao Lee Institute (TDLI), Shanghai Jiao Tong University, Shanghai, China
\item[$^{13}$] Universidad Nacional Aut\'onoma de M\'exico (UNAM), Instituto de Ciencias Nucleares, M\'exico, D.F., M\'exico
\item[$^{14}$] University of Hawai'i at Manoa, Department of Physics and Astronomy, Honolulu, USA
\item[$^{15}$] Karlsruhe Institute of Technology (KIT), Institute of Experimental Particle Physics (ETP), Karlsruhe, Germany
\item[$^{16}$] Laborat\'orio de Instrumenta\c{c}\~ao e F\'\i{}sica Experimental de Part\'\i{}culas (LIP), Lisboa, Portugal
\item[$^{17}$] Fluka collaboration
\end{description}

\vspace{-1ex}
\footnotesize
\section*{Acknowledgments}
This research has been funded by the Federal Ministry of Education and Research of Germany and the state of North-Rhine Westphalia as part of the Lamarr-Institute for Machine Learning and Artificial Intelligence.

\end{document}